\documentclass[superscriptaddress,reprint]{revtex4-1} 

\usepackage{graphicx}
\usepackage{graphics}
\usepackage{amssymb}
\usepackage{amsmath,bm}
\usepackage{float}
\usepackage[usenames,dvipsnames]{color}
\usepackage[export]{adjustbox}
\usepackage{hyperref}
\hypersetup{
    colorlinks=true,
    linkcolor=blue,
    citecolor=blue,
    filecolor=cyan,      
    urlcolor=cyan,
}
\begin{document}

\title{Dynamics of trapped atoms around an optical nanofiber probed through polarimetry}

\author{Pablo Solano}
\affiliation{Joint Quantum Institute, Department of Physics, University of Maryland and NIST, College Park, MD
20742, USA.}

\author{Fredrik K. Fatemi}
\affiliation{Army Research Laboratory, Adelphi, MD 20783, USA.}

\author{Luis A. Orozco}
\affiliation{Joint Quantum Institute, Department of Physics, University of Maryland and NIST, College Park, MD
20742, USA.}

\author{S. L. Rolston}
\affiliation{Joint Quantum Institute, Department of Physics, University of Maryland and NIST, College Park, MD
20742, USA.}


\begin{abstract}
The evanescent field outside an optical nanofiber (ONF) can create optical traps for neutral atoms. We present a non-destructive method to characterize such trapping potentials. An off-resonance linearly polarized probe beam that propagates through the ONF experiences a slow axis of polarization produced by trapped atoms on opposite sides along the ONF. The transverse atomic motion is imprinted onto the probe polarization through the changing atomic index of of refraction. By applying a transient impulse, we measure a time-dependent polarization rotation of the probe beam that provides both a rapid and non-destructive measurement of the optical trapping frequencies.

\end{abstract}

\maketitle

Nano-optical waveguides allow efficient ways to couple trapped atoms to propagating photons, a crucial element in the development of quantum technologies \cite{Thompson2013,Goban2014,Goban2015,Hood2016}. Optical nanofibers (ONF) \cite{Morrissey2013} have shown to be a particularly versatile platform in this context by enabling quantum memories \cite{Gouraud2015,Sayrin2015,Jones2015,Kumar2015}, switches \cite{OShea2013,Shomroni2014}, diodes \cite{Sayrin2015a}, and reflectors \cite{Corzo2016,Sorensen2016}. These examples show integration of photonic and atomic systems.

An ONF consists of single-mode optical fiber heated and pulled to create a tapered profile. The tapers can adiabatically guide the propagating light in and out of a sub-wavelength diameter waist with less than 0.1\% loss~\cite{Hoffman2014a}. 
Because the nanofiber radius is smaller than the wavelength of the propagating mode, most of the field is outside its dielectric body as an evanescent field \cite{LeKien2004}. This field allows  coupling of atoms near the ONF surface to the guided mode. The tight confinement of the propagating mode enables significant atom-light coupling. 

The large spatial gradient of the evanescent field enables an optical dipole trap for atoms with two different wavelengths of light, one detuned above atomic resonance (blue-detuned) to repel the atoms from the surface, and the other detuned below resonance (red-detuned) for confinement.  Such traps are an effective tool to confine atoms close the the ONF waveguide for millisecond time-scales with low optical powers ($\approx$5 mW), creating a robust platform for coupling propagating photons to atoms \cite{Vetsch2010,Goban2012,Reitz2013,Beguin2014a,Kato2015}.

\begin{figure}[htbp]
\centering
\includegraphics[width=0.9\linewidth]{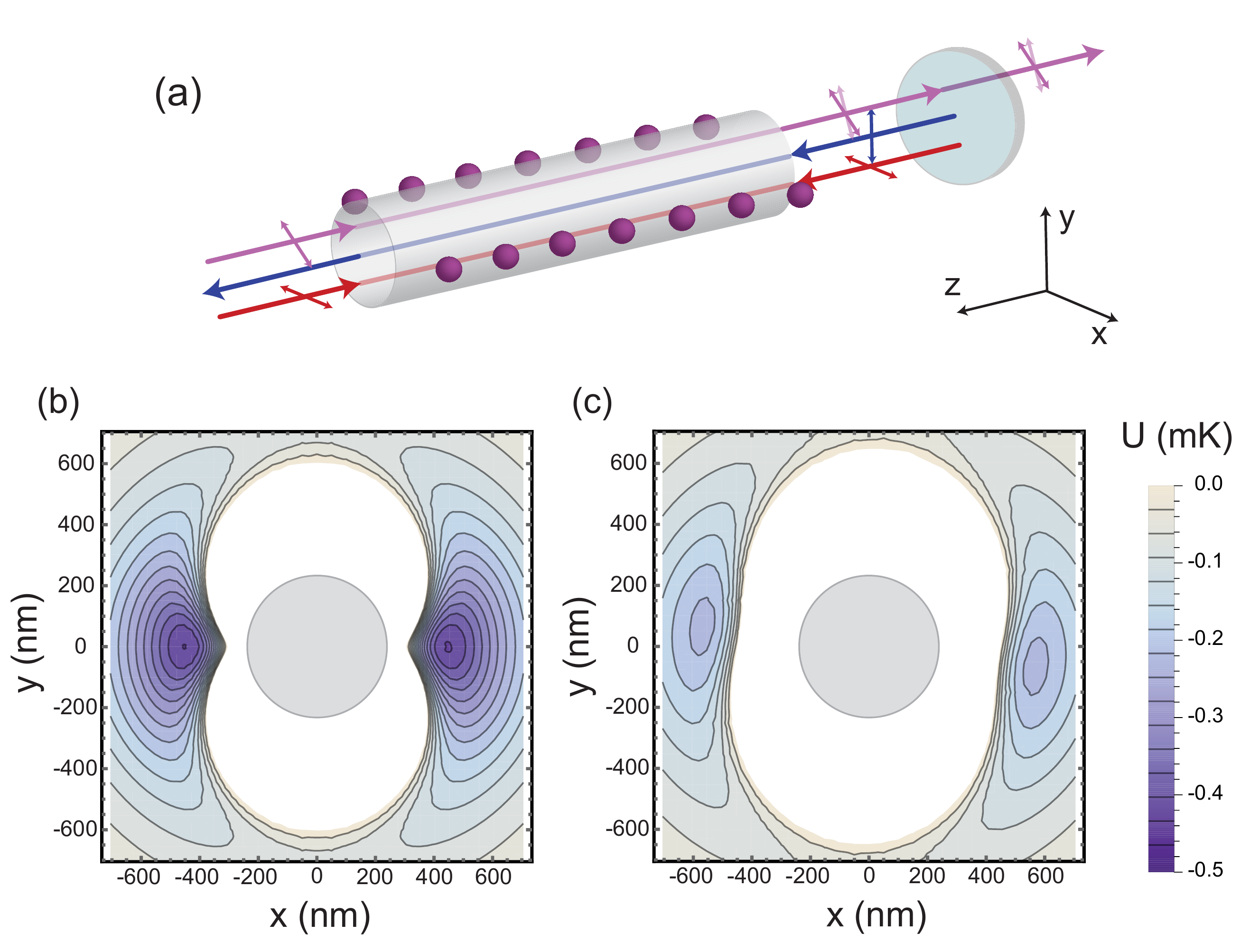}
\caption{(a) Schematic of the experimental setup showing the two one-dimensional array of atoms. An off-resonance probe beam propagates through the sample with linear polarization rotated by 45$^{\circ}$.  (b) Transversal view of a trapping potential, with 1 mW of power in each red-detuned beam and 3 mW of blue detuned propagating through a 235-nm radius ONF waist. (c) Modification of the trapping potential in (b) after turning on a probe beam with 70 nW of power and 200 MHz detuned to the blue of atomic resonance.}
\label{fig:1}
\end{figure}

A typical ONF dipole trap, with retro-reflection of the red-detuned light, creates two one-dimensional arrays of atoms on each side of the ONF, sketched in Fig. \ref{fig:1} (a). Characterizing the atom number and trap characteristics is necessary for future applications of this platform. The number of trapped atoms can be measured on resonance \cite{Vetsch2010} or off resonance \cite{Beguin2014a,Qi2016}, \textit{i.e.} destructive and dispersive measurements, respectively. Parametric heating to find vibrational frequencies has also been applied to ONFs~\cite{Vetsch2010a}, but is destructive and is a serial measurement for finding the trap frequencies.

In this letter we present a method to non-destructively characterize the trapping potential of an ONF dipole trap. We propagate a weak, off-resonance probe beam through the ONF that is linearly polarized and tilted 45$^{\circ}$ relative to the azimuthal axis defined by the trapping potential. The probe experiences a modified refractive index with a fast axis and a slow axis due to the presence of trapped atoms. This effective birefringence rotates the polarization of the probe as a function of the position of the atoms. Turning on the probe beam imparts a momentum kick to the trapped atoms so that they oscillate at the radial and azimuthal trapping frequencies. Detecting the time-dependent polarization change of the probe gives us a direct and non-destructive measurement of the motion and transverse frequencies of the trapping potential. By probing the atomic motion directly, the spectrum of the system response can be analyzed in a single time-domain measurement up to the bandwidth of the detection.

Because the evanescent field decay constant is proportional to its wavelength, the red (blue) detuned light creates a longer (shorter) range attractive (repulsive) potential. Combining both red and blue detuned light, the atoms experience a potential energy minimum a fraction of a wavelength away from the ONF surface. This two-color dipole trap provides radial confinement for the atoms. Two counter-propagating red-detuned beams in a standing-wave configuration provide confinement along the optical nanofiber in a one-dimensional lattice. Azimuthal confinement is achieved by correctly choosing the polarization of the trapping beams. At the ONF waist, linearly-polarized light becomes quasi-linearly polarized, breaking the azimuthal symmetry of the intensity profile of the propagating field. Aligning the polarization axis of the red-detuned beam orthogonal to the blue detuned one provides azimuthal confinement for the atoms (See Fig. \ref{fig:1} (a) and (b)).

We create a dipole trap for $^{87}$Rb atoms with a 235-nm radius ONF waist by coupling two counter-propagating red-detuned beams (1064 nm) in a standing wave configuration and one blue-detuned beam (750 nm). The dominant resonances for Rb are at 780 nm (D2 line)  and 795 nm (D1 line). We typically use 1 mW of power for each red-detuned beam, and 3 mW for the blue-detuned beam. Fig.~\ref{fig:1}(b) shows this configuration, which produces a trapping potential with a depth of about  500 $\mu$K. Here, and throughout the paper, we consider only the atomic scalar polarizability for the calculations of the trapping potentials.

We image the light scattered from the nanofiber to characterize the polarization of the laser beams at the ONF waist~\cite{Hoffman2015}. Because Rayleigh scattering preserves the polarization of the field, with the help of a linear polarizer in front of the camera we determine the polarization of the propagating field. The polarization can be controlled by wave plates at the input of the ONF. Each laser beam has to be characterized and controlled independently, since inherent stress in the ONF creates a birefringent medium that affects each wavelength differently.

A magneto-optical trap (MOT) loads cold $^{87}$Rb atoms into our ONF dipole trap in a vacuum chamber kept at lower than $10^{-9}$ Torr. We further cool the atoms by increasing the detuning of the MOT beams for 90 ms. We then turn off the magnetic field gradient to create optical molasses for 1 ms. The atoms are typically at 15 $\mu$K when we let them fall into the dipole trap. Because of the tight confinement of the trap, the atoms are expected to be in a collisional blockade regime. This leads to a binary loading with one or zero atoms per trapping site. We typically trap a few hundred atoms for trapping lifetimes of the order of 10 ms. The trapped atoms are in an statistical mixture of $m_F$ Zeeman sub-levels.

We send an off-resonant beam, detuned 200 MHz to the blue of the the $F=2\rightarrow F'=3$ transition of the D2 line, through the ONF to probe the trapped atoms. We align its polarization to be 45$^{\circ}$ from the trapping beams when there are no atoms present. The projection of the transverse polarization component along the axis defined by the trapped atoms experiences a modified refractive index while the orthogonal component, which does not interact with the atoms, propagates unaltered. The motion of trapped atoms in the transverse plane of the nanofiber will change this birefringence as a function of time, producing a dynamical polarization rotation of the probe beam.  Motion along the fiber axis (z-direction) is likely to be only weakly coupled to the probe and would not produce significant polarization rotation.

 \begin{figure}[htbp]
\centering
\includegraphics[width=0.9\linewidth]{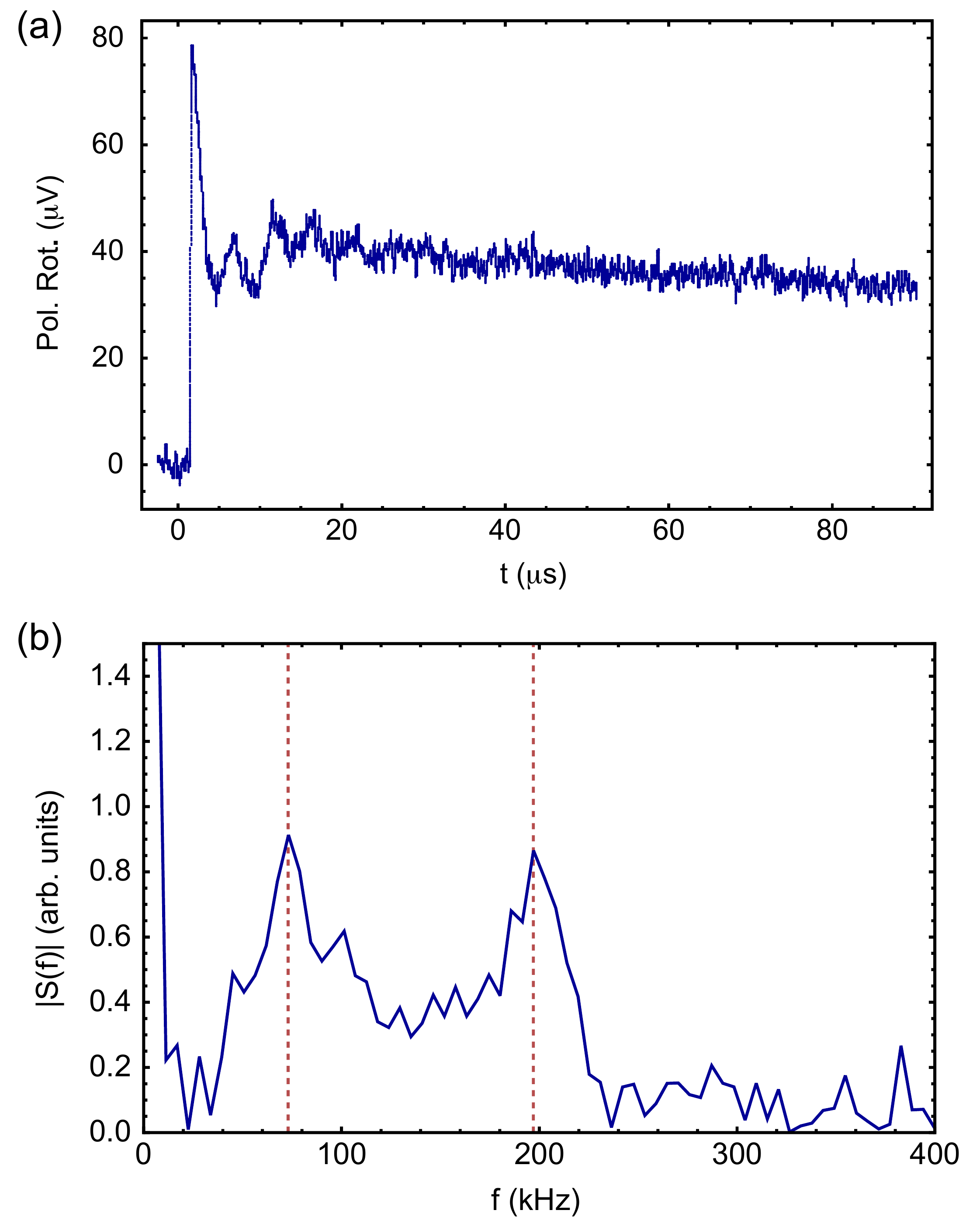}
\caption{(a) Polarization rotation of the probe beam (in units of measured voltage) as a function of time. The sudden spike in the signal denotes the probe turning on. (b) Power spectrum from the Fourier transform of the oscillations in (a). The two distinct peaks (at 73 $\pm 3$ kHz and 197 $\pm 2$ kHz), correspond to the radial and azimuthal trapping frequencies respectively, marked with red dashed lines.}
\label{fig:2}
\end{figure}

Because of the significant atom-light coupling provided by the tight mode area, more than a few tens of nW of probe power will perturb the trap near resonance. We use 70 nW of probe power, enough to imprint a momentum kick in the atoms to start their motion, but too weak to excite the atoms out of the trap. Fig. \ref{fig:1} (c) shows the effect of the probe beam on the trapping potential. 

The polarization rotation of such a low probe power is detected by heterodyne measurements by mixing the probe with a local oscillator (LO) with a 1 MHz relative frequency shift. We typically use 9 mW of power for the LO beam. After the probe goes through the ONF it is combined with the LO using a 50/50 beam splitter. We use one of the output paths for detection. Its polarization components are separated by a Wollaston prism and sent to a 4 MHz bandwidth balanced photodetector. The 1 MHz beat note between the probe and the LO is mixed down to DC. This allows us to use the LO as gain for the probe, and directly detect the probe polarization rotation as a function of time with a bandwidth higher than the expected trap frequencies.

Figure \ref{fig:2} (a) shows a typical signal of the polarization rotation of the probe. Although the signal is visible in single-shot, the data is averaged to improve the signal to noise ratio by a factor of 10.  The original data was acquired with a 2-ns bin width, and the plot is a 400-ns moving average for visualization purposes. The detector polarizations are set such that when there are no trapped atoms the measured output voltage is zero. However the zero voltage at time $t=0$ in the plot is produced only by the LO (probe beam off). The probe field turns on at 2 $\mu$s. The signal can be decomposed in two time regimes: a short time regime where we observe oscillations due to the atoms moving back and forth in the trapping potential; and a long time regime where the oscillations vanish but the non-zero signal shows the presence of atoms in the trap. The sharp initial peak comes from atoms starting their motion closer to the ONF surface, where they interact more strongly with the probe beam, producing a larger signal. The decoherence of the oscillations comes from the large anharmonicity of the trapping potential and the thermal motion of the trapped atoms. The long timescale slope is the lifetime of the trap. In this case the characteristic decay time is $370\pm 3$ $\mu$s, where the error represents the standard error of the fit. The lifetime is degraded by more than an order of magnitude when the probe beam is kept on. A small fraction of the probe beam gets absorbed by the trapped atoms and results in losses as the trapping potential becomes shallower (see Figs. \ref{fig:1} (b) and (c) with the depth scale).

The temporal response and initial oscillations in Fig. \ref{fig:2} (a) encode information about transverse trapping frequencies. By taking a discrete Fourier transform of the data (after the probe turns on) we obtain the resonance frequencies of the oscillating atoms. Fig. \ref{fig:2} (b) shows the power spectrum of the signal. We observe two distinct peaks at $\nu_{\phi}=73\pm 3 $ kHz and $\nu_{r}=197\pm 2 $ kHz, corresponding to the azimuthal and radial frequencies of the trap. The uncertainties in the mean are calculated from the full width at half maximum of the peak over the signal to noise ratio \cite{Clairon1991}. The width of the spectral peaks and damping of the time-domain oscillations arise from the dephasing of the atoms due to the strong anharmonicity of the trap. As an approximation, we can model the problem as a damped harmonic oscillator. The fit to a Lorentzian line shape shows a linewidth of $\gamma_{\phi}=64\pm 8$ kHz $\gamma_{r}=47\pm 6$ kHz respectively, where the errors are the standard errors of the fit. This represents a decay time of the oscillations of around 20 $\mu$s, enough to measure trapping potentials of more than 50 kHz. The observation of oscillations from the azimuthal motion of the atoms depends on the alignment of the probe polarization to within few degrees. On the other hand, the detection of oscillation from radial motion of the atoms is more robust under misalignments.

We can compare the measured frequencies in Fig. \ref{fig:2} (b) to a numerical calculation. Taking the second derivative of the trapping potential shown in Fig.\ref{fig:1} (c) and knowing the atomic mass $m$ we can calculate the expected trapping frequencies as $\nu_{i}=\sqrt{\frac{1}{2\pi m}\partial^2 U/\partial x_i^2}$, where the index $i$ denotes the radial or azimuthal direction in cylindrical coordinates. For the experimental parameters listed in this paper, which produce Fig. \ref{fig:2} (c), we find that $\nu_{\phi}=70\pm 4$ kHz and $\nu_{r}=195\pm 6$ kHz. The frequencies are extracted by fit an harmonic potential to the bottom of the calculated potential and extracting the corresponding trapping frequency for each spatial direction. The errors represent the sensitivity of the simulation to a 5\% variation of the experimental parameters, these parameters being the four lasers beams power (two red-detuned, a blue-detuned and the probe), and the four polarization angles (three relative angles). We assume that the polarizations are perfectly linearly-polarized, which is in general not true, but greatly reduces the number of free parameters in the simulation. The theoretical results are 2\% above and 7\% below the measured values for the azimuthal and radial frequencies respectively. The measured signal is in good agreement with the expected result within the experimental uncertainties.

\begin{figure}[htbp]
\centering
\includegraphics[width=0.9\linewidth]{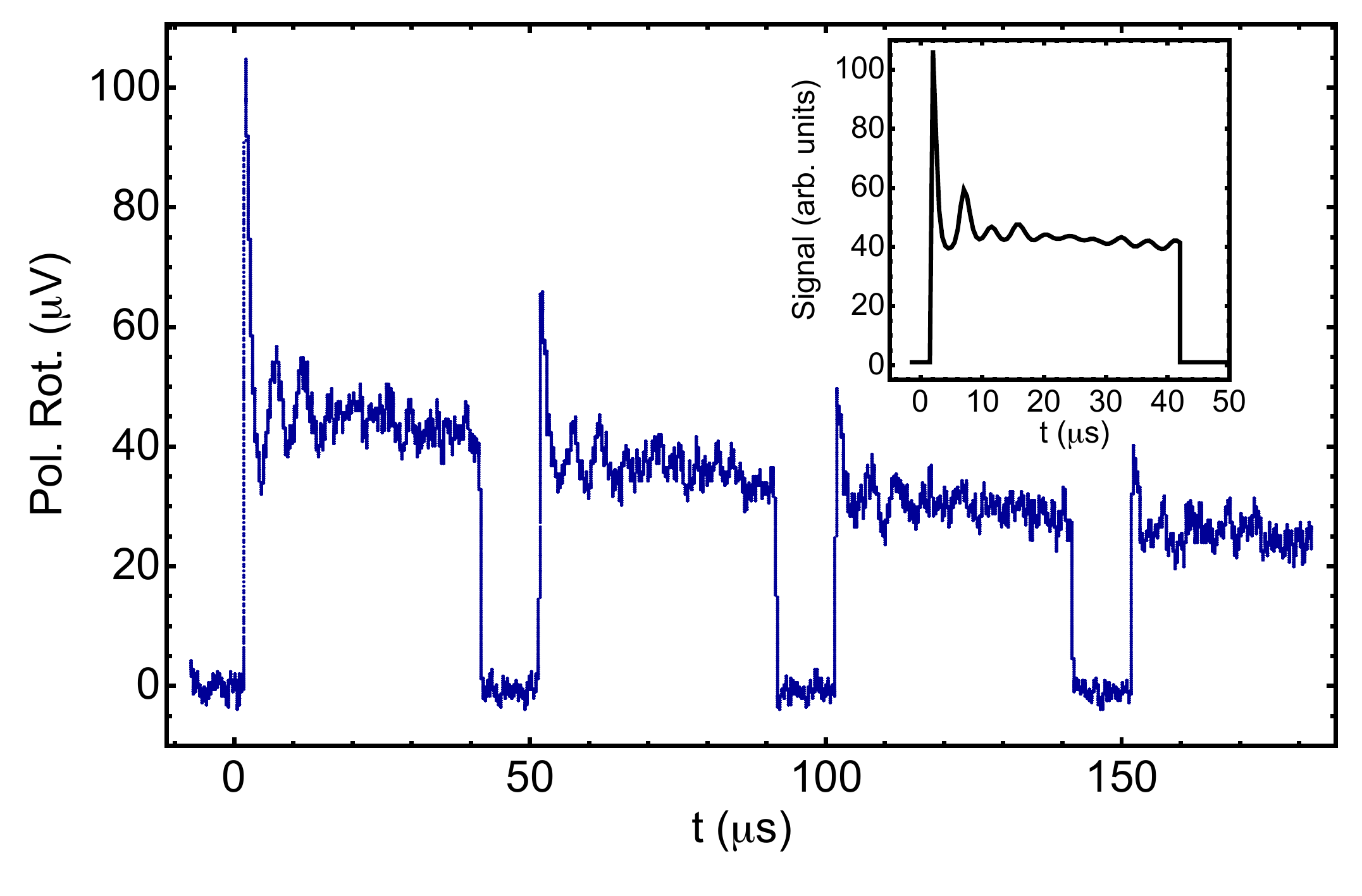}
\caption{Polarization rotation of the probe beam (in units of measured voltage) as a function of time, for a set of four 40 $\mu$s probe pulses. The repeatability of the process shows the non-destructive feature of the measurement technique. The inset shows a Monte Carlo simulation of the signal for radial oscillations only. The simulation considers an ensemble of atoms oscillating in the potential shown in Fig. \ref{fig:1} (c) from different starting positions and a decay of $265$ $\mu$s.}
\label{fig:3}
\end{figure}

The non-destructive feature of this method is further tested by probing the trapped atoms more than once while they still are in the trap. Fig. \ref{fig:3} shows the polarization rotation as a function of time for a probe beam that turns on and off four times. We see that the first pulse is enough to extract the oscillation frequency of the atoms before it decreases. Consecutively the probe turns off and on again, after 10 $\mu$s, reproducing the same oscillatory signal but with smaller amplitude. This process can be repeated as long there are enough atoms in the trap to produce a detectable signal. The signal from the four pulses shown in Fig. \ref{fig:3} has an over-all slope corresponding to a trapping lifetime of $265\pm 1$ $\mu$s. This is almost 30\% shorter lifetime compared to keeping the probe beam constantly on (as in Fig. \ref{fig:2} (a)), because the momentum kick of suddenly turning the probe beam on and off can induce atom loss. However, the dispersive measurement is non-destructive enough to test the characteristics of the trap while leaving a significant amount of atoms for further experimentation. The inset of Fig. \ref{fig:3} shows a numerical simulation of the detected signal for only radial oscillations (uncoupled motion). Using the simulated trapping potential (Fig \ref{fig:1} (c)) we calculate the motion of a set of 500 atoms randomly positioned with a flat distribution of $\pm 75$ nm centered at $80$ nm towards the ONF from the potential minimum. The trajectories of the atoms, computed and averaged, give an effective trajectory. The signal is proportional to the dynamical change of the coupling into the ONF of an atom following such an effective trajectory. The displacement of the center of the distribution of the initial atomic positions takes into account the displacement of the center of the trap when the probe beam is turned on. The parameters for the simulation are empirically found within a experimentally realistic range. This simple model captures the qualitative behavior of the detected signal.

Although the probe beam modifies the potential landscape felt by the atoms, the good agreement between the measurements and the simulations allows us to extract the trapping potential without the modification due to the probe beam. In our case we obtain $\nu_{\phi}=178.3$ KHz and $\nu_{\phi}=252.2$ KHz from the potential shown in Fig. \ref{fig:1} (b). Moreover, by optimizing the photodetection, a weaker probe beam could be used to minimally perturb the trapping potential. In this configuration another pulsed beam can rapidly imprint a momentum kick to the atoms, so they start oscillating in phase. Colder atoms might also help to establish longer coherence time for the oscillations, since the trapping potential approximates to an harmonic trap around its minimum. The measured signal increases linearly with the number of trapped atoms. A more efficient loading of the trap may increase the number of atoms and the amplitude of the signal.

We have shown how a polarimetric measurement of an off-resonance probe beam can be used for rapid and non-destructive characterization of the trapping potential of a two-color ONF-based dipole trap. This technique can be easily implemented in any ONF-based dipole trap experiment, allowing a shot-to-shot measurement of the trapping potential before performing further experiments in the same experimental sequence, an advantage over other configurations of optical dipole traps. The results are in good agreement with theoretical predictions, showing an understanding of the variables involved in the problem. This points to different strategies to improve the technique in the future. We expect that non-destructive and fast-readout characterization of local potential experienced by trapped atoms near dielectric surfaces to become standard tools in the growing field of interfacing nano-photonic platforms to cold atoms.

\section*{Acknowledgments}
This work has been supported by National Science Foundation of the United States (NSF) (PHY-1307416); NSF Physics Frontier Center at the Joint quantum Institute (PHY-1430094).
\bibliography{NanofibersRef}

\begin{thebibliography}{25}%
\makeatletter
\providecommand \@ifxundefined [1]{%
 \@ifx{#1\undefined}
}%
\providecommand \@ifnum [1]{%
 \ifnum #1\expandafter \@firstoftwo
 \else \expandafter \@secondoftwo
 \fi
}%
\providecommand \@ifx [1]{%
 \ifx #1\expandafter \@firstoftwo
 \else \expandafter \@secondoftwo
 \fi
}%
\providecommand \natexlab [1]{#1}%
\providecommand \enquote  [1]{``#1''}%
\providecommand \bibnamefont  [1]{#1}%
\providecommand \bibfnamefont [1]{#1}%
\providecommand \citenamefont [1]{#1}%
\providecommand \href@noop [0]{\@secondoftwo}%
\providecommand \href [0]{\begingroup \@sanitize@url \@href}%
\providecommand \@href[1]{\@@startlink{#1}\@@href}%
\providecommand \@@href[1]{\endgroup#1\@@endlink}%
\providecommand \@sanitize@url [0]{\catcode `\\12\catcode `\$12\catcode
  `\&12\catcode `\#12\catcode `\^12\catcode `\_12\catcode `\%12\relax}%
\providecommand \@@startlink[1]{}%
\providecommand \@@endlink[0]{}%
\providecommand \url  [0]{\begingroup\@sanitize@url \@url }%
\providecommand \@url [1]{\endgroup\@href {#1}{\urlprefix }}%
\providecommand \urlprefix  [0]{URL }%
\providecommand \Eprint [0]{\href }%
\providecommand \doibase [0]{http://dx.doi.org/}%
\providecommand \selectlanguage [0]{\@gobble}%
\providecommand \bibinfo  [0]{\@secondoftwo}%
\providecommand \bibfield  [0]{\@secondoftwo}%
\providecommand \translation [1]{[#1]}%
\providecommand \BibitemOpen [0]{}%
\providecommand \bibitemStop [0]{}%
\providecommand \bibitemNoStop [0]{.\EOS\space}%
\providecommand \EOS [0]{\spacefactor3000\relax}%
\providecommand \BibitemShut  [1]{\csname bibitem#1\endcsname}%
\let\auto@bib@innerbib\@empty
\bibitem [{\citenamefont {Thompson}\ \emph {et~al.}(2013)\citenamefont
  {Thompson}, \citenamefont {Tiecke}, \citenamefont {de~Leon}, \citenamefont
  {Feist}, \citenamefont {Akimov}, \citenamefont {Gullans}, \citenamefont
  {Zibrov}, \citenamefont {Vuleti\'{c}},\ and\ \citenamefont
  {Lukin}}]{Thompson2013}%
  \BibitemOpen
  \bibfield  {author} {\bibinfo {author} {\bibfnamefont {J.~D.}\ \bibnamefont
  {Thompson}}, \bibinfo {author} {\bibfnamefont {T.~G.}\ \bibnamefont
  {Tiecke}}, \bibinfo {author} {\bibfnamefont {N.~P.}\ \bibnamefont {de~Leon}},
  \bibinfo {author} {\bibfnamefont {J.}~\bibnamefont {Feist}}, \bibinfo
  {author} {\bibfnamefont {A.~V.}\ \bibnamefont {Akimov}}, \bibinfo {author}
  {\bibfnamefont {M.}~\bibnamefont {Gullans}}, \bibinfo {author} {\bibfnamefont
  {A.~S.}\ \bibnamefont {Zibrov}}, \bibinfo {author} {\bibfnamefont
  {V.}~\bibnamefont {Vuleti\'{c}}}, \ and\ \bibinfo {author} {\bibfnamefont
  {M.~D.}\ \bibnamefont {Lukin}},\ }\href {\doibase 10.1126/science.1237125}
  {\bibfield  {journal} {\bibinfo  {journal} {Science}\ }\textbf {\bibinfo
  {volume} {340}},\ \bibinfo {pages} {1202} (\bibinfo {year}
  {2013})}\BibitemShut {NoStop}%
\bibitem [{\citenamefont {Goban}\ \emph {et~al.}(2014)\citenamefont {Goban},
  \citenamefont {Hung}, \citenamefont {Yu}, \citenamefont {Hood}, \citenamefont
  {Muniz}, \citenamefont {Lee}, \citenamefont {Martin}, \citenamefont
  {McClung}, \citenamefont {Choi}, \citenamefont {Chang}, \citenamefont
  {Painter},\ and\ \citenamefont {Kimble}}]{Goban2014}%
  \BibitemOpen
  \bibfield  {author} {\bibinfo {author} {\bibfnamefont {A.}~\bibnamefont
  {Goban}}, \bibinfo {author} {\bibfnamefont {C.-L.}\ \bibnamefont {Hung}},
  \bibinfo {author} {\bibfnamefont {S.-P.}\ \bibnamefont {Yu}}, \bibinfo
  {author} {\bibfnamefont {J.}~\bibnamefont {Hood}}, \bibinfo {author}
  {\bibfnamefont {J.}~\bibnamefont {Muniz}}, \bibinfo {author} {\bibfnamefont
  {J.}~\bibnamefont {Lee}}, \bibinfo {author} {\bibfnamefont {M.}~\bibnamefont
  {Martin}}, \bibinfo {author} {\bibfnamefont {A.}~\bibnamefont {McClung}},
  \bibinfo {author} {\bibfnamefont {K.}~\bibnamefont {Choi}}, \bibinfo {author}
  {\bibfnamefont {D.}~\bibnamefont {Chang}}, \bibinfo {author} {\bibfnamefont
  {O.}~\bibnamefont {Painter}}, \ and\ \bibinfo {author} {\bibfnamefont
  {H.}~\bibnamefont {Kimble}},\ }\href {\doibase 10.1038/ncomms4808} {\bibfield
   {journal} {\bibinfo  {journal} {Nat. Commun.}\ }\textbf {\bibinfo {volume}
  {5}} (\bibinfo {year} {2014}),\ 10.1038/ncomms4808}\BibitemShut {NoStop}%
\bibitem [{\citenamefont {Goban}\ \emph {et~al.}(2015)\citenamefont {Goban},
  \citenamefont {Hung}, \citenamefont {Hood}, \citenamefont {Yu}, \citenamefont
  {Muniz}, \citenamefont {Painter},\ and\ \citenamefont {Kimble}}]{Goban2015}%
  \BibitemOpen
  \bibfield  {author} {\bibinfo {author} {\bibfnamefont {A.}~\bibnamefont
  {Goban}}, \bibinfo {author} {\bibfnamefont {C.-L.}\ \bibnamefont {Hung}},
  \bibinfo {author} {\bibfnamefont {J.~D.}\ \bibnamefont {Hood}}, \bibinfo
  {author} {\bibfnamefont {S.-P.}\ \bibnamefont {Yu}}, \bibinfo {author}
  {\bibfnamefont {J.~A.}\ \bibnamefont {Muniz}}, \bibinfo {author}
  {\bibfnamefont {O.}~\bibnamefont {Painter}}, \ and\ \bibinfo {author}
  {\bibfnamefont {H.~J.}\ \bibnamefont {Kimble}},\ }\href {\doibase
  10.1103/PhysRevLett.115.063601} {\bibfield  {journal} {\bibinfo  {journal}
  {Phys. Rev. Lett.}\ }\textbf {\bibinfo {volume} {115}},\ \bibinfo {pages}
  {063601} (\bibinfo {year} {2015})}\BibitemShut {NoStop}%
\bibitem [{\citenamefont {Hood}\ \emph {et~al.}(2016)\citenamefont {Hood},
  \citenamefont {Goban}, \citenamefont {Asenjo-Garcia}, \citenamefont {Lu},
  \citenamefont {Yu}, \citenamefont {Chang},\ and\ \citenamefont
  {Kimble}}]{Hood2016}%
  \BibitemOpen
  \bibfield  {author} {\bibinfo {author} {\bibfnamefont {J.~D.}\ \bibnamefont
  {Hood}}, \bibinfo {author} {\bibfnamefont {A.}~\bibnamefont {Goban}},
  \bibinfo {author} {\bibfnamefont {A.}~\bibnamefont {Asenjo-Garcia}}, \bibinfo
  {author} {\bibfnamefont {M.}~\bibnamefont {Lu}}, \bibinfo {author}
  {\bibfnamefont {S.-P.}\ \bibnamefont {Yu}}, \bibinfo {author} {\bibfnamefont
  {D.~E.}\ \bibnamefont {Chang}}, \ and\ \bibinfo {author} {\bibfnamefont
  {H.~J.}\ \bibnamefont {Kimble}},\ }\href {\doibase 10.1073/pnas.1603788113}
  {\bibfield  {journal} {\bibinfo  {journal} {Proceedings of the National
  Academy of Sciences}\ }\textbf {\bibinfo {volume} {113}},\ \bibinfo {pages}
  {10507} (\bibinfo {year} {2016})},\ \Eprint
  {http://arxiv.org/abs/http://www.pnas.org/content/113/38/10507.full.pdf}
  {http://www.pnas.org/content/113/38/10507.full.pdf} \BibitemShut {NoStop}%
\bibitem [{\citenamefont {Morrissey}\ \emph {et~al.}(2013)\citenamefont
  {Morrissey}, \citenamefont {Deasy}, \citenamefont {Frawley}, \citenamefont
  {Kumar}, \citenamefont {Prel}, \citenamefont {Russell}, \citenamefont
  {Truong},\ and\ \citenamefont {{Nic~Chormaic}}}]{Morrissey2013}%
  \BibitemOpen
  \bibfield  {author} {\bibinfo {author} {\bibfnamefont {M.~J.}\ \bibnamefont
  {Morrissey}}, \bibinfo {author} {\bibfnamefont {K.}~\bibnamefont {Deasy}},
  \bibinfo {author} {\bibfnamefont {M.}~\bibnamefont {Frawley}}, \bibinfo
  {author} {\bibfnamefont {R.}~\bibnamefont {Kumar}}, \bibinfo {author}
  {\bibfnamefont {E.}~\bibnamefont {Prel}}, \bibinfo {author} {\bibfnamefont
  {L.}~\bibnamefont {Russell}}, \bibinfo {author} {\bibfnamefont {V.~G.}\
  \bibnamefont {Truong}}, \ and\ \bibinfo {author} {\bibfnamefont
  {S.}~\bibnamefont {{Nic~Chormaic}}},\ }\href {\doibase 10.3390/s130810449}
  {\bibfield  {journal} {\bibinfo  {journal} {Sensors (Basel).}\ }\textbf
  {\bibinfo {volume} {13}},\ \bibinfo {pages} {10449} (\bibinfo {year}
  {2013})}\BibitemShut {NoStop}%
\bibitem [{\citenamefont {Gouraud}\ \emph {et~al.}(2015)\citenamefont
  {Gouraud}, \citenamefont {Maxein}, \citenamefont {Nicolas}, \citenamefont
  {Morin},\ and\ \citenamefont {Laurat}}]{Gouraud2015}%
  \BibitemOpen
  \bibfield  {author} {\bibinfo {author} {\bibfnamefont {B.}~\bibnamefont
  {Gouraud}}, \bibinfo {author} {\bibfnamefont {D.}~\bibnamefont {Maxein}},
  \bibinfo {author} {\bibfnamefont {A.}~\bibnamefont {Nicolas}}, \bibinfo
  {author} {\bibfnamefont {O.}~\bibnamefont {Morin}}, \ and\ \bibinfo {author}
  {\bibfnamefont {J.}~\bibnamefont {Laurat}},\ }\href {\doibase
  10.1103/PhysRevLett.114.180503} {\bibfield  {journal} {\bibinfo  {journal}
  {Phys. Rev. Lett.}\ }\textbf {\bibinfo {volume} {114}},\ \bibinfo {pages}
  {180503} (\bibinfo {year} {2015})}\BibitemShut {NoStop}%
\bibitem [{\citenamefont {Sayrin}\ \emph
  {et~al.}(2015{\natexlab{a}})\citenamefont {Sayrin}, \citenamefont {Clausen},
  \citenamefont {Albrecht}, \citenamefont {Schneeweiss},\ and\ \citenamefont
  {Rauschenbeutel}}]{Sayrin2015}%
  \BibitemOpen
  \bibfield  {author} {\bibinfo {author} {\bibfnamefont {C.}~\bibnamefont
  {Sayrin}}, \bibinfo {author} {\bibfnamefont {C.}~\bibnamefont {Clausen}},
  \bibinfo {author} {\bibfnamefont {B.}~\bibnamefont {Albrecht}}, \bibinfo
  {author} {\bibfnamefont {P.}~\bibnamefont {Schneeweiss}}, \ and\ \bibinfo
  {author} {\bibfnamefont {A.}~\bibnamefont {Rauschenbeutel}},\ }\href
  {\doibase 10.1364/OPTICA.2.000353} {\bibfield  {journal} {\bibinfo  {journal}
  {Optica}\ }\textbf {\bibinfo {volume} {2}},\ \bibinfo {pages} {353} (\bibinfo
  {year} {2015}{\natexlab{a}})}\BibitemShut {NoStop}%
\bibitem [{\citenamefont {Jones}\ \emph {et~al.}(2015)\citenamefont {Jones},
  \citenamefont {Franson},\ and\ \citenamefont {Pittman}}]{Jones2015}%
  \BibitemOpen
  \bibfield  {author} {\bibinfo {author} {\bibfnamefont {D.~E.}\ \bibnamefont
  {Jones}}, \bibinfo {author} {\bibfnamefont {J.~D.}\ \bibnamefont {Franson}},
  \ and\ \bibinfo {author} {\bibfnamefont {T.~B.}\ \bibnamefont {Pittman}},\
  }\href {\doibase 10.1103/PhysRevA.92.043806} {\bibfield  {journal} {\bibinfo
  {journal} {Phys. Rev. A}\ }\textbf {\bibinfo {volume} {92}},\ \bibinfo
  {pages} {043806} (\bibinfo {year} {2015})}\BibitemShut {NoStop}%
\bibitem [{\citenamefont {Kumar}\ \emph {et~al.}(2015)\citenamefont {Kumar},
  \citenamefont {Gokhroo}, \citenamefont {Deasy}, \citenamefont {Maimaiti},
  \citenamefont {Frawley}, \citenamefont {Phelan},\ and\ \citenamefont
  {{Nic~Chormaic}}}]{Kumar2015}%
  \BibitemOpen
  \bibfield  {author} {\bibinfo {author} {\bibfnamefont {R.}~\bibnamefont
  {Kumar}}, \bibinfo {author} {\bibfnamefont {V.}~\bibnamefont {Gokhroo}},
  \bibinfo {author} {\bibfnamefont {K.}~\bibnamefont {Deasy}}, \bibinfo
  {author} {\bibfnamefont {A.}~\bibnamefont {Maimaiti}}, \bibinfo {author}
  {\bibfnamefont {M.~C.}\ \bibnamefont {Frawley}}, \bibinfo {author}
  {\bibfnamefont {C.}~\bibnamefont {Phelan}}, \ and\ \bibinfo {author}
  {\bibfnamefont {S.}~\bibnamefont {{Nic~Chormaic}}},\ }\href {\doibase
  10.1088/1367-2630/17/1/013026} {\bibfield  {journal} {\bibinfo  {journal}
  {New J. Phys.}\ }\textbf {\bibinfo {volume} {17}},\ \bibinfo {pages} {013026}
  (\bibinfo {year} {2015})}\BibitemShut {NoStop}%
\bibitem [{\citenamefont {O'Shea}\ \emph {et~al.}(2013)\citenamefont {O'Shea},
  \citenamefont {Junge}, \citenamefont {Volz},\ and\ \citenamefont
  {Rauschenbeutel}}]{OShea2013}%
  \BibitemOpen
  \bibfield  {author} {\bibinfo {author} {\bibfnamefont {D.}~\bibnamefont
  {O'Shea}}, \bibinfo {author} {\bibfnamefont {C.}~\bibnamefont {Junge}},
  \bibinfo {author} {\bibfnamefont {J.}~\bibnamefont {Volz}}, \ and\ \bibinfo
  {author} {\bibfnamefont {A.}~\bibnamefont {Rauschenbeutel}},\ }\href
  {\doibase 10.1103/PhysRevLett.111.193601} {\bibfield  {journal} {\bibinfo
  {journal} {Phys. Rev. Lett.}\ }\textbf {\bibinfo {volume} {111}},\ \bibinfo
  {pages} {193601} (\bibinfo {year} {2013})}\BibitemShut {NoStop}%
\bibitem [{\citenamefont {Shomroni}\ \emph {et~al.}(2014)\citenamefont
  {Shomroni}, \citenamefont {Rosenblum}, \citenamefont {Lovsky}, \citenamefont
  {Bechler}, \citenamefont {Guendelman},\ and\ \citenamefont
  {Dayan}}]{Shomroni2014}%
  \BibitemOpen
  \bibfield  {author} {\bibinfo {author} {\bibfnamefont {I.}~\bibnamefont
  {Shomroni}}, \bibinfo {author} {\bibfnamefont {S.}~\bibnamefont {Rosenblum}},
  \bibinfo {author} {\bibfnamefont {Y.}~\bibnamefont {Lovsky}}, \bibinfo
  {author} {\bibfnamefont {O.}~\bibnamefont {Bechler}}, \bibinfo {author}
  {\bibfnamefont {G.}~\bibnamefont {Guendelman}}, \ and\ \bibinfo {author}
  {\bibfnamefont {B.}~\bibnamefont {Dayan}},\ }\href {\doibase
  10.1126/science.1254699} {\bibfield  {journal} {\bibinfo  {journal}
  {Science}\ }\textbf {\bibinfo {volume} {345}},\ \bibinfo {pages} {903}
  (\bibinfo {year} {2014})},\ \Eprint
  {http://arxiv.org/abs/http://science.sciencemag.org/content/345/6199/903.full.pdf}
  {http://science.sciencemag.org/content/345/6199/903.full.pdf} \BibitemShut
  {NoStop}%
\bibitem [{\citenamefont {Sayrin}\ \emph
  {et~al.}(2015{\natexlab{b}})\citenamefont {Sayrin}, \citenamefont {Junge},
  \citenamefont {Mitsch}, \citenamefont {Albrecht}, \citenamefont {O'Shea},
  \citenamefont {Schneeweiss}, \citenamefont {Volz},\ and\ \citenamefont
  {Rauschenbeutel}}]{Sayrin2015a}%
  \BibitemOpen
  \bibfield  {author} {\bibinfo {author} {\bibfnamefont {C.}~\bibnamefont
  {Sayrin}}, \bibinfo {author} {\bibfnamefont {C.}~\bibnamefont {Junge}},
  \bibinfo {author} {\bibfnamefont {R.}~\bibnamefont {Mitsch}}, \bibinfo
  {author} {\bibfnamefont {B.}~\bibnamefont {Albrecht}}, \bibinfo {author}
  {\bibfnamefont {D.}~\bibnamefont {O'Shea}}, \bibinfo {author} {\bibfnamefont
  {P.}~\bibnamefont {Schneeweiss}}, \bibinfo {author} {\bibfnamefont
  {J.}~\bibnamefont {Volz}}, \ and\ \bibinfo {author} {\bibfnamefont
  {A.}~\bibnamefont {Rauschenbeutel}},\ }\href {\doibase
  10.1103/PhysRevX.5.041036} {\bibfield  {journal} {\bibinfo  {journal} {Phys.
  Rev. X}\ }\textbf {\bibinfo {volume} {5}},\ \bibinfo {pages} {041036}
  (\bibinfo {year} {2015}{\natexlab{b}})}\BibitemShut {NoStop}%
\bibitem [{\citenamefont {Corzo}\ \emph {et~al.}(2016)\citenamefont {Corzo},
  \citenamefont {Gouraud}, \citenamefont {Chandra}, \citenamefont {Goban},
  \citenamefont {Sheremet}, \citenamefont {Kupriyanov},\ and\ \citenamefont
  {Laurat}}]{Corzo2016}%
  \BibitemOpen
  \bibfield  {author} {\bibinfo {author} {\bibfnamefont {N.~V.}\ \bibnamefont
  {Corzo}}, \bibinfo {author} {\bibfnamefont {B.}~\bibnamefont {Gouraud}},
  \bibinfo {author} {\bibfnamefont {A.}~\bibnamefont {Chandra}}, \bibinfo
  {author} {\bibfnamefont {A.}~\bibnamefont {Goban}}, \bibinfo {author}
  {\bibfnamefont {A.~S.}\ \bibnamefont {Sheremet}}, \bibinfo {author}
  {\bibfnamefont {D.}~\bibnamefont {Kupriyanov}}, \ and\ \bibinfo {author}
  {\bibfnamefont {J.}~\bibnamefont {Laurat}},\ }\href {\doibase
  10.1103/PhysRevLett.117.133603} {\bibfield  {journal} {\bibinfo  {journal}
  {Phys. Rev. Lett.}\ }\textbf {\bibinfo {volume} {117}},\ \bibinfo {pages}
  {133603} (\bibinfo {year} {2016})}\BibitemShut {NoStop}%
\bibitem [{\citenamefont {S\o{}rensen}\ \emph {et~al.}(2016)\citenamefont
  {S\o{}rensen}, \citenamefont {B\'eguin}, \citenamefont {Kluge}, \citenamefont
  {Iakoupov}, \citenamefont {S\o{}rensen}, \citenamefont {M\"uller},
  \citenamefont {Polzik},\ and\ \citenamefont {Appel}}]{Sorensen2016}%
  \BibitemOpen
  \bibfield  {author} {\bibinfo {author} {\bibfnamefont {H.~L.}\ \bibnamefont
  {S\o{}rensen}}, \bibinfo {author} {\bibfnamefont {J.-B.}\ \bibnamefont
  {B\'eguin}}, \bibinfo {author} {\bibfnamefont {K.~W.}\ \bibnamefont {Kluge}},
  \bibinfo {author} {\bibfnamefont {I.}~\bibnamefont {Iakoupov}}, \bibinfo
  {author} {\bibfnamefont {A.~S.}\ \bibnamefont {S\o{}rensen}}, \bibinfo
  {author} {\bibfnamefont {J.~H.}\ \bibnamefont {M\"uller}}, \bibinfo {author}
  {\bibfnamefont {E.~S.}\ \bibnamefont {Polzik}}, \ and\ \bibinfo {author}
  {\bibfnamefont {J.}~\bibnamefont {Appel}},\ }\href {\doibase
  10.1103/PhysRevLett.117.133604} {\bibfield  {journal} {\bibinfo  {journal}
  {Phys. Rev. Lett.}\ }\textbf {\bibinfo {volume} {117}},\ \bibinfo {pages}
  {133604} (\bibinfo {year} {2016})}\BibitemShut {NoStop}%
\bibitem [{\citenamefont {Hoffman}\ \emph {et~al.}(2014)\citenamefont
  {Hoffman}, \citenamefont {Ravets}, \citenamefont {Grover}, \citenamefont
  {Solano}, \citenamefont {Kordell}, \citenamefont {Wong-Campos}, \citenamefont
  {Orozco},\ and\ \citenamefont {Rolston}}]{Hoffman2014a}%
  \BibitemOpen
  \bibfield  {author} {\bibinfo {author} {\bibfnamefont {J.~E.}\ \bibnamefont
  {Hoffman}}, \bibinfo {author} {\bibfnamefont {S.}~\bibnamefont {Ravets}},
  \bibinfo {author} {\bibfnamefont {J.~A.}\ \bibnamefont {Grover}}, \bibinfo
  {author} {\bibfnamefont {P.}~\bibnamefont {Solano}}, \bibinfo {author}
  {\bibfnamefont {P.~R.}\ \bibnamefont {Kordell}}, \bibinfo {author}
  {\bibfnamefont {J.~D.}\ \bibnamefont {Wong-Campos}}, \bibinfo {author}
  {\bibfnamefont {L.~A.}\ \bibnamefont {Orozco}}, \ and\ \bibinfo {author}
  {\bibfnamefont {S.~L.}\ \bibnamefont {Rolston}},\ }\href {\doibase
  10.1063/1.4879799} {\bibfield  {journal} {\bibinfo  {journal} {AIP Adv.}\
  }\textbf {\bibinfo {volume} {4}},\ \bibinfo {pages} {067124} (\bibinfo {year}
  {2014})}\BibitemShut {NoStop}%
\bibitem [{\citenamefont {Le~Kien}\ \emph {et~al.}(2004)\citenamefont
  {Le~Kien}, \citenamefont {Balykin},\ and\ \citenamefont
  {Hakuta}}]{LeKien2004}%
  \BibitemOpen
  \bibfield  {author} {\bibinfo {author} {\bibfnamefont {F.}~\bibnamefont
  {Le~Kien}}, \bibinfo {author} {\bibfnamefont {V.~I.}\ \bibnamefont
  {Balykin}}, \ and\ \bibinfo {author} {\bibfnamefont {K.}~\bibnamefont
  {Hakuta}},\ }\href {\doibase 10.1103/PhysRevA.70.063403} {\bibfield
  {journal} {\bibinfo  {journal} {Phys. Rev. A}\ }\textbf {\bibinfo {volume}
  {70}},\ \bibinfo {pages} {063403} (\bibinfo {year} {2004})}\BibitemShut
  {NoStop}%
\bibitem [{\citenamefont {Vetsch}\ \emph {et~al.}(2010)\citenamefont {Vetsch},
  \citenamefont {Reitz}, \citenamefont {Sagu\'e}, \citenamefont {Schmidt},
  \citenamefont {Dawkins},\ and\ \citenamefont {Rauschenbeutel}}]{Vetsch2010}%
  \BibitemOpen
  \bibfield  {author} {\bibinfo {author} {\bibfnamefont {E.}~\bibnamefont
  {Vetsch}}, \bibinfo {author} {\bibfnamefont {D.}~\bibnamefont {Reitz}},
  \bibinfo {author} {\bibfnamefont {G.}~\bibnamefont {Sagu\'e}}, \bibinfo
  {author} {\bibfnamefont {R.}~\bibnamefont {Schmidt}}, \bibinfo {author}
  {\bibfnamefont {S.~T.}\ \bibnamefont {Dawkins}}, \ and\ \bibinfo {author}
  {\bibfnamefont {A.}~\bibnamefont {Rauschenbeutel}},\ }\href {\doibase
  10.1103/PhysRevLett.104.203603} {\bibfield  {journal} {\bibinfo  {journal}
  {Phys. Rev. Lett.}\ }\textbf {\bibinfo {volume} {104}},\ \bibinfo {pages}
  {203603} (\bibinfo {year} {2010})}\BibitemShut {NoStop}%
\bibitem [{\citenamefont {Goban}\ \emph {et~al.}(2012)\citenamefont {Goban},
  \citenamefont {Choi}, \citenamefont {Alton}, \citenamefont {Ding},
  \citenamefont {Lacro\^ute}, \citenamefont {Pototschnig}, \citenamefont
  {Thiele}, \citenamefont {Stern},\ and\ \citenamefont {Kimble}}]{Goban2012}%
  \BibitemOpen
  \bibfield  {author} {\bibinfo {author} {\bibfnamefont {A.}~\bibnamefont
  {Goban}}, \bibinfo {author} {\bibfnamefont {K.~S.}\ \bibnamefont {Choi}},
  \bibinfo {author} {\bibfnamefont {D.~J.}\ \bibnamefont {Alton}}, \bibinfo
  {author} {\bibfnamefont {D.}~\bibnamefont {Ding}}, \bibinfo {author}
  {\bibfnamefont {C.}~\bibnamefont {Lacro\^ute}}, \bibinfo {author}
  {\bibfnamefont {M.}~\bibnamefont {Pototschnig}}, \bibinfo {author}
  {\bibfnamefont {T.}~\bibnamefont {Thiele}}, \bibinfo {author} {\bibfnamefont
  {N.~P.}\ \bibnamefont {Stern}}, \ and\ \bibinfo {author} {\bibfnamefont
  {H.~J.}\ \bibnamefont {Kimble}},\ }\href {\doibase
  10.1103/PhysRevLett.109.033603} {\bibfield  {journal} {\bibinfo  {journal}
  {Phys. Rev. Lett.}\ }\textbf {\bibinfo {volume} {109}},\ \bibinfo {pages}
  {033603} (\bibinfo {year} {2012})}\BibitemShut {NoStop}%
\bibitem [{\citenamefont {Reitz}\ \emph {et~al.}(2013)\citenamefont {Reitz},
  \citenamefont {Sayrin}, \citenamefont {Mitsch}, \citenamefont {Schneeweiss},\
  and\ \citenamefont {Rauschenbeutel}}]{Reitz2013}%
  \BibitemOpen
  \bibfield  {author} {\bibinfo {author} {\bibfnamefont {D.}~\bibnamefont
  {Reitz}}, \bibinfo {author} {\bibfnamefont {C.}~\bibnamefont {Sayrin}},
  \bibinfo {author} {\bibfnamefont {R.}~\bibnamefont {Mitsch}}, \bibinfo
  {author} {\bibfnamefont {P.}~\bibnamefont {Schneeweiss}}, \ and\ \bibinfo
  {author} {\bibfnamefont {A.}~\bibnamefont {Rauschenbeutel}},\ }\href
  {\doibase 10.1103/PhysRevLett.110.243603} {\bibfield  {journal} {\bibinfo
  {journal} {Phys. Rev. Lett.}\ }\textbf {\bibinfo {volume} {110}},\ \bibinfo
  {pages} {243603} (\bibinfo {year} {2013})}\BibitemShut {NoStop}%
\bibitem [{\citenamefont {B\'{e}guin}\ \emph {et~al.}(2014)\citenamefont
  {B\'{e}guin}, \citenamefont {Bookjans}, \citenamefont {Christensen},
  \citenamefont {S\o{}rensen}, \citenamefont {M\"{u}ller}, \citenamefont
  {Polzik},\ and\ \citenamefont {Appel}}]{Beguin2014a}%
  \BibitemOpen
  \bibfield  {author} {\bibinfo {author} {\bibfnamefont {J.-B.}\ \bibnamefont
  {B\'{e}guin}}, \bibinfo {author} {\bibfnamefont {E.}~\bibnamefont
  {Bookjans}}, \bibinfo {author} {\bibfnamefont {S.}~\bibnamefont
  {Christensen}}, \bibinfo {author} {\bibfnamefont {H.}~\bibnamefont
  {S\o{}rensen}}, \bibinfo {author} {\bibfnamefont {J.}~\bibnamefont
  {M\"{u}ller}}, \bibinfo {author} {\bibfnamefont {E.}~\bibnamefont {Polzik}},
  \ and\ \bibinfo {author} {\bibfnamefont {J.}~\bibnamefont {Appel}},\ }\href
  {\doibase 10.1103/PhysRevLett.113.263603} {\bibfield  {journal} {\bibinfo
  {journal} {Phys. Rev. Lett.}\ }\textbf {\bibinfo {volume} {113}},\ \bibinfo
  {pages} {263603} (\bibinfo {year} {2014})}\BibitemShut {NoStop}%
\bibitem [{\citenamefont {Kato}\ and\ \citenamefont {Aoki}(2015)}]{Kato2015}%
  \BibitemOpen
  \bibfield  {author} {\bibinfo {author} {\bibfnamefont {S.}~\bibnamefont
  {Kato}}\ and\ \bibinfo {author} {\bibfnamefont {T.}~\bibnamefont {Aoki}},\
  }\href {\doibase 10.1103/PhysRevLett.115.093603} {\bibfield  {journal}
  {\bibinfo  {journal} {Phys. Rev. Lett.}\ }\textbf {\bibinfo {volume} {115}},\
  \bibinfo {pages} {093603} (\bibinfo {year} {2015})}\BibitemShut {NoStop}%
\bibitem [{\citenamefont {Qi}\ \emph {et~al.}(2016)\citenamefont {Qi},
  \citenamefont {Baragiola}, \citenamefont {Jessen},\ and\ \citenamefont
  {Deutsch}}]{Qi2016}%
  \BibitemOpen
  \bibfield  {author} {\bibinfo {author} {\bibfnamefont {X.}~\bibnamefont
  {Qi}}, \bibinfo {author} {\bibfnamefont {B.~Q.}\ \bibnamefont {Baragiola}},
  \bibinfo {author} {\bibfnamefont {P.~S.}\ \bibnamefont {Jessen}}, \ and\
  \bibinfo {author} {\bibfnamefont {I.~H.}\ \bibnamefont {Deutsch}},\ }\href
  {\doibase 10.1103/PhysRevA.93.023817} {\bibfield  {journal} {\bibinfo
  {journal} {Phys. Rev. A}\ }\textbf {\bibinfo {volume} {93}},\ \bibinfo
  {pages} {023817} (\bibinfo {year} {2016})}\BibitemShut {NoStop}%
\bibitem [{\citenamefont {Vetsch}(2010)}]{Vetsch2010a}%
  \BibitemOpen
  \bibfield  {author} {\bibinfo {author} {\bibfnamefont {E.}~\bibnamefont
  {Vetsch}},\ }\emph {\bibinfo {title} {{Optical Interface Based on a Nanofiber
  Atom-Trap}}},\ \href@noop {} {Ph.D. thesis},\ \bibinfo  {school} {University
  of Mainz} (\bibinfo {year} {2010})\BibitemShut {NoStop}%
\bibitem [{\citenamefont {Hoffman}\ \emph {et~al.}(2015)\citenamefont
  {Hoffman}, \citenamefont {Fatemi}, \citenamefont {Beadie}, \citenamefont
  {Rolston},\ and\ \citenamefont {Orozco}}]{Hoffman2015}%
  \BibitemOpen
  \bibfield  {author} {\bibinfo {author} {\bibfnamefont {J.~E.}\ \bibnamefont
  {Hoffman}}, \bibinfo {author} {\bibfnamefont {F.~K.}\ \bibnamefont {Fatemi}},
  \bibinfo {author} {\bibfnamefont {G.}~\bibnamefont {Beadie}}, \bibinfo
  {author} {\bibfnamefont {S.~L.}\ \bibnamefont {Rolston}}, \ and\ \bibinfo
  {author} {\bibfnamefont {L.~A.}\ \bibnamefont {Orozco}},\ }\href {\doibase
  10.1364/OPTICA.2.000416} {\bibfield  {journal} {\bibinfo  {journal} {Optica}\
  }\textbf {\bibinfo {volume} {2}},\ \bibinfo {pages} {416} (\bibinfo {year}
  {2015})}\BibitemShut {NoStop}%
\bibitem [{\citenamefont {Clairon}\ \emph {et~al.}(1991)\citenamefont
  {Clairon}, \citenamefont {Salomon}, \citenamefont {Guellati},\ and\
  \citenamefont {Phillips}}]{Clairon1991}%
  \BibitemOpen
  \bibfield  {author} {\bibinfo {author} {\bibfnamefont {A.}~\bibnamefont
  {Clairon}}, \bibinfo {author} {\bibfnamefont {C.}~\bibnamefont {Salomon}},
  \bibinfo {author} {\bibfnamefont {S.}~\bibnamefont {Guellati}}, \ and\
  \bibinfo {author} {\bibfnamefont {W.~D.}\ \bibnamefont {Phillips}},\ }\href
  {http://stacks.iop.org/0295-5075/16/i=2/a=008} {\bibfield  {journal}
  {\bibinfo  {journal} {EPL (Europhysics Letters)}\ }\textbf {\bibinfo {volume}
  {16}},\ \bibinfo {pages} {165} (\bibinfo {year} {1991})}\BibitemShut
  {NoStop}%
\end{thebibliography}%

\end{document}